\documentclass{article}
\usepackage{INTERSPEECH2021}
\usepackage{spconf}
\usepackage{amsmath,amssymb,multirow, cite}
\usepackage{url}
\usepackage{siunitx}
\usepackage[nolist]{acronym}
\usepackage{booktabs}

\usepackage{multirow}
\usepackage{subfigure}
\newacro{CRNN}[CRNN]{Convolutional Recurrent Neural Network}
\newacro{CNN}[CNN]{Convolutional Neural Network}
\newacro{WER}[WER]{Word Error Rate}
\newacro{WA}[WA]{Word Accuracy}
\newacro{ASR}[ASR]{Automatic Speech Recognition}
\newacro{WPE}[WPE]{Weighted Prediction Error}
\newacro{DAB}[DAB]{Deep Ad-hoc Beamforming}
\newacro{EV}[EV]{Envelope Variance}
\newacro{DNN}[DNN]{Deep Neural Network}
\newacro{ASR}[ASR]{Automatic Speech Recognition}
\newacro{SNR}[SNR]{Signal-to-Noise Ratio}
\ninept

\title{Learning to Rank Microphones for Distant Speech Recognition}

\name{Samuele Cornell$^1$, Alessio Brutti$^2$, Marco Matassoni$^2$, Stefano Squartini$^1$}
\address{
  $^1$Università Politecnica delle Marche\\
  $^2$Fondazione Bruno Kessler}

\begin{document}

\maketitle

%Most of current channel selection techniques rely on signal-level features in an attempt to find the cleanest channel. However, these

%metrics do not always correlate with the recognition performance.

\begin{abstract}
Fully exploiting ad-hoc microphone networks for distant speech recognition is still an open issue. 
Empirical evidence shows that being able to select the best microphone leads to significant improvements in recognition without any additional effort on front-end processing.
Current channel selection techniques either rely on signal, decoder or posterior-based features. 
Signal-based features are inexpensive to compute but do not always correlate with recognition performance. Instead decoder and posterior-based features exhibit better correlation but require substantial computational resources. 

In this work, we tackle the channel selection problem by proposing MicRank, a learning to rank framework where a neural network is trained to rank the available channels using directly the recognition performance on the training set. The proposed approach is agnostic with respect to the array geometry and type of recognition back-end. We investigate different learning to rank strategies using a synthetic dataset developed on purpose and the CHi\-ME-6 data. 
Results show that the proposed approach is able to considerably improve over previous selection techniques, reaching comparable and in some instances better performance than oracle signal-based measures. 

%Empirical evidence shows that being able to pick the best microphone can potentially lead to very significant WER reduction without any further effort in terms of speech processing, speech enhancement or acoustic modeling. Current solutions typically rely on some signal-level features (e.g. SNR, distance) in an attempt to detect the cleanest channel. However, these physical metrics do not always correlate well with the behavior of the ASR, as they do not account for the actual ASR acoustic and language models. %One of the limitations of existing signal processing methods is that they just aim for the cleanest signal without accounting for the actual ASR acoustic modeling.

% Samuele: .
%to predict the recognition performance on a given channel using the actual word error rate on the training set. The proposed approach is agnostic with respect to the array geometry and type of recognition back-end employed as it is trained to rank the channels according to the back-end errors.   

%The resulting model is agnostic of the physics of the acoustic propagation and of the acquisition setup, while it is expected to learn what the ASR likes the most in terms of acoustic properties of the signal. 

%Samuele: the proposed approach is agnostic with respect to the array geometry and type of ASR back-end employed as it is trained to select the channel which will minimize its error.  

 %to WER reductions with respect to the baseline and other channel selection methods.
\end{abstract}
\begin{keywords}
speech recognition, channel selection, learning to rank, array signal processing
\end{keywords}

\section{Introduction}
\label{sec:intro}

Nowadays, many application scenarios envision the presence of multiple heterogeneous recording devices. Examples are meeting scenarios \cite{Yoshioka_2019} or multi-party conversations such as in CHi\-ME-6 Challenge~\cite{CHiME6}. However, distant \ac{ASR} in presence of ad-hoc microphone networks is still an open issue and the potential of fusing information from multiple devices towards the common goal of reducing the~\ac{WER} is still not fully exploited.

Audio signals captured by different microphones can be suitably combined at front-end level by using beamforming techniques \cite{bai2013acoustic, Beamformit, erdogan2016improved, heymann2016neural, boeddeker2018front, zhang2021deep,  TAC2020}. However, most of these approaches \cite{bai2013acoustic, Beamformit, erdogan2016improved, heymann2016neural } are designed for microphone array applications and do not perform well in ad-hoc microphone scenarios where sensors can be far from each other. 
Few exceptions are \cite{Yoshioka_2019, zhang2021deep, TAC2020} in which ad-hoc microphone networks are explicitly considered in the design of the method. %These combination methods, while effective, have the drawback that they can require significant computational resources especially for \ac{DNN}-based beamforming methods. 
%, either fine-tuning or end-to-end training with the ASR back-end can be necessary in order to provide substantial WER reduction \cite{}. 

%If asynchronous devices are considered this combination is further complicated. Recently, in~\cite{Yoshioka_2019} the synchronization problem is tackled by performing a continuous alignment of the audio stream while the actual channel combination is based on Recognizer Output Voting Error Reduction (ROVER)~\cite{ROVER} on multiple combined microphone subsets

%Moreover, DNN-based beamforming methods significant recognition gain for DNN-based either fine-tuning or, directly, end-to-end training is usually required \cite{}. 

%Empirical experiments on the data used for the CHi\-ME-5 and CHi\-ME-6 challenges shows that this simple approach can provide large improvements in terms of \ac{WER} reduction [REF?].

Another intriguing approach, is to pick up, for each utterance, the best channel without any further processing, or, in alternative, sorting the channels and choosing a promising subset before applying signal-based combination methods or Recognizer Output Voting Error Reduction (ROVER)~\cite{ROVER}.
This channel selection problem has been addressed in the past either using signal-based hand-crafted features \cite{kumatani2011channel, WOLF_EV_2014, Guerrero2018}, decoder-based \cite{Obuchi-2006, wolfel2007channel} and posterior-based features \cite{Xiong2018}. 
Among the most representative past studies on automatic channel selection, \cite{WOLF_EV_2014} (and previous works from the same authors) investigated both signal-based and decoder-based measures, as well as different strategies for their combination. It was found that \ac{EV}, despite being signal-based, represents one the most effective channel selection strategy thanks to its ability to detect the reduced dynamic ranges introduced by reverberation.
More recently, in \cite{Guerrero2018}, another signal-based method relying on Cepstral Distance (CD) was proposed. 
The main advantage of these signal-based methods is that they are inexpensive to compute with respect to decoder-based measures which require full decoding of all channels.
Another option is the posterior-based channel selection method proposed in \cite{Xiong2018}
in which microphones are selected using an entropy measure of posterior probabilities produced by an Acoustic Model (AM) trained on clean speech. %The basic assumption is that channels with low entropy will eventually lead to lower error rate, since they better match the clean input signals used to train this acoustic model. 
While less expensive than decoder-based methods, as it requires only an AM forward pass for each channel, it assumes that a matched AM trained on clean speech is available. 

In this paper, we propose MicRank, an alternative, fully neural approach for channel selection. MicRank is agnostic with respect to the properties of the acoustic environment, recording set up and \ac{ASR} back-end.
Borrowing from information retrieval \cite{GUO_survey_2019}, we formulate the channel selection task as a learning to rank (LTR) problem where a DNN is trained to rank microphones based on the errors obtained with the \ac{ASR} back-end on a training set. 
Within this framework, we explore different loss functions and training strategies by performing experiments on a purposely developed synthetic dataset and CHi\-ME-6. 
We show that MicRank considerably outperforms several previously proposed channel selection methods and even, in some instances, signal-based oracle measures. Importantly, this is achieved with remarkably lower computational requirements compared to decoder and posterior-based approaches. 
Our source code is made open source at \url{github.com/popcornell/MicRank}.

This paper is organized as follows.  Section~\ref{sec:learningtorank} presents the learning to rank paradigm and how it can be adapted to address the channel selection problem. Section~\ref{sec:experiments} describes the experimental set-up, including datasets, baseline methods and neural architectures. Following, in Section~\ref{sec:results} we discuss our experimental results and, in Section~\ref{sec:conclusions}, we draw conclusions and outline possible future research directions. %concludes the paper with the final remarks and plan for future activities.
\vspace{-0.2cm}
\section{Learning to Rank for Channel Selection}
\label{sec:learningtorank}
The problem of selecting the best channel among a set of available ones can be best formulated as a ranking problem. In fact, predicting an absolute quality metric (e.g. \ac{WER}, \ac{SNR} etc.) for each channel is not necessary as what matter most is relative performance: for a given utterance we want to find the best channel within the available ones, whatever its absolute quality metric is. This requires the model to learn, either implicitly or explicitly, to rank the channels. %and a set of microphone channels it is desired to find the best one compared to the other available channels. This, either implicitly or explicitly, requires the model to learn to rank the channels from the best performing one to the worst one. To reach this goal, a suitable learning-to-rank training strategy can be adopted. A significant advantage of this framework is that the inference phase can still be treated as a classical single channel regression problem: the network predicts, for each channel, a relevance score which can be then used to rank the channels. This allows the selection algorithm to be invariant to the number and to the particular permutation of the channels.

%As a matter of facts we are not interested in predicting the channel quality in an absolute sense but we just want to pick what the ASR likes the most among the available channels. Note that while we have to adopt particular training strategies to achieve this goal, in inference we are basically dealing with a traditional single channel regression problem.

Learning to rank is an established framework in the field of information retrieval. Therefore its formulation has to be revised and adapted to channel selection for ASR, in particular for what concerns the relevance of observed samples. 
In principle the ranking approaches we propose can be used to rank the channels with respect to any metric. Since in this work the ultimate goal is ASR, training labels are derived directly from \ac{WER} or \ac{WA} obtained by the \ac{ASR} back-end on the training material. 
% say smthing about ROVER ? 

% Therefore we have to revise the formulation in order to make it fit our ASR application context, in particular for what concerns the relevance of the observed samples. Since our goal is to improve the ASR performance, we derive the training labels from the \ac{WER} or the~\ac{WA} obtained on the training material.

%It is worth mentioning that, in order to be able to generalize to different configurations, models must be single-channel. Considering multiple channel in input would work only if the same geometric configuration is used. Conversely, multiple channels can be combined in training by evaluating multi-channel losses.

%\subfigure[Point Wise]{
%\includegraphics[width=.2\textwidth]{pointwise.png}
%\label{fig:point}
%}
%\subfigure[RankNet]{
%\includegraphics[width=.35\textwidth]{ranknet.png}
%\label{fig:rank}
%}
%\subfigure[ListNet]{

%\includegraphics[width=.39\textwidth]{ListNet.pdf}
%\label{fig:list}
%}

\subsection{Ranking Strategies and Losses}
\begin{figure*}[htbp]
\centering
\includegraphics[width=15cm]{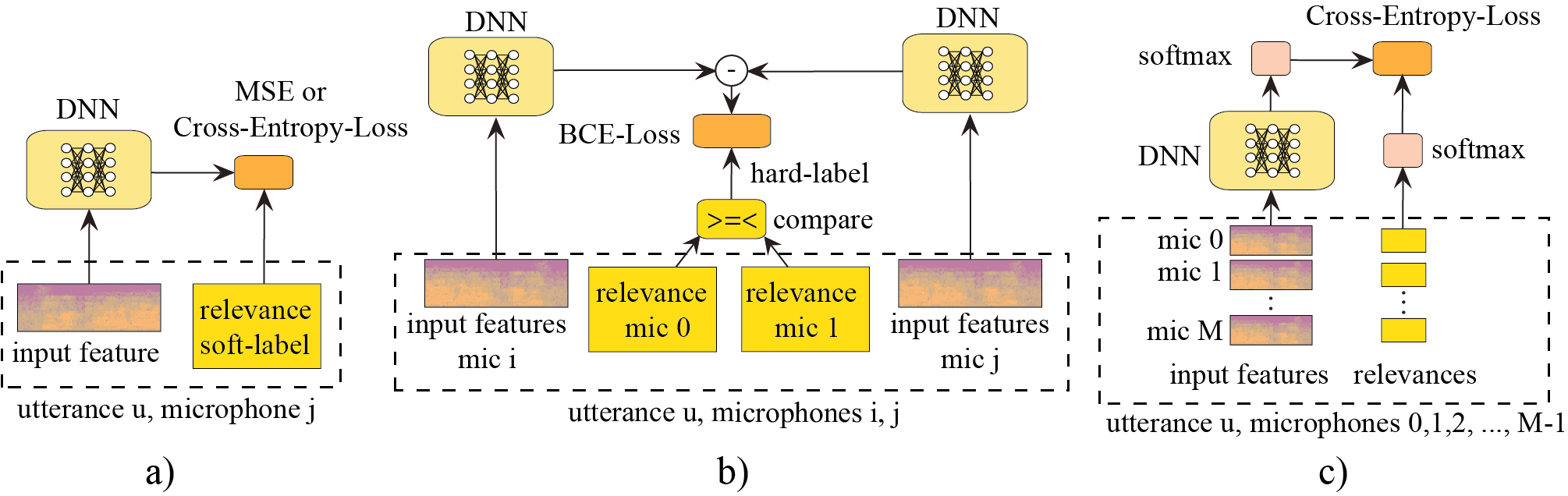}

\caption{Training strategies: a) point-wise training; b) pair-wise training with RankNet; c) list-wise training with ListNet.}
\label{fig:training}
\vspace{-0.5cm}
\end{figure*}
%Let us assume that overall $U$ utterances are recorded by $M$ microphones. For each utterance $u$ ($u=0,\dots,U-1$), given the set of observation features $\bm{x}_u=(x_{u,0}, x_{u,1}, \dots, x_{u,M-1})$ extracted from $M$ microphone signals and a ranking order (or relevance in information retrieval) $w_{u,i}$ for each $x_{u,i}$, our goal is to identify a function $f(x_{u,i})$ that generates the same ranking order: if $w_{u,i} > w_{u,j}$ then $f(x_{u,i}) > f(x_{u,j})$.
Let us assume that $U$ utterances are recorded by $M$ microphones. For each utterance $u$ ($u=0,\dots,U-1$), given the observation feature vector $\bm{x}_{u,i}$ for the $i$-th microphone ($i=0,\dots,M-1$) and a ranking order (or relevance in information retrieval) $w_{u,i}$, our goal is to define a function $f(\bm{x}_{u,i})$ that generates the same ranking order: if $w_{u,i} > w_{u,j}$ then $f(\bm{x}_{u,i}) > f(\bm{x}_{u,j})$. 
In the following we describe different training strategies to achieve this goal, graphically depicted in Fig.~\ref{fig:training}.
\subsubsection{Point-wise training}
The most straightforward approach to rank the channels is to employ a model trained on each single channel individually to predict its relevance.
In this method, given a set of %utterance features $\bm{x}_{u,i}$ and relevance $w_{u,i}$ 
training pairs $(\bm{x}_{u,i},w_{u,i})$ for each utterance and microphone
 the network is trained to minimize a cross-entropy loss: 
\begin{equation}
%\mathcal{L}_{\mbox{\tiny{XCE}}-1}=\sum_{\left\{\bm{x}_{u,i},w_{u,i}\right\}\in \mathcal{X}} w_{u,i} \log\left[\sigma(f(\bm{x}_{u,i}))\right],
\mathcal{L}_{\mbox{\tiny{XCE}}}^{point}=\sum_{u=0}^{U-1}\sum_{i=0}^{M-1} w_{u,i}\log\left[\sigma(f(\bm{x}_{u,i}))\right],
\end{equation}
where $\sigma(\cdot)$ is the sigmoid operator. In this case, the relevance label $0\leq w_j \leq 1$ is a soft label, representing the quality of the speech signal in an absolute term. ~\ac{WA} for example, and any other bounded metric can be used straightforwardly. A clipping or normalization strategy instead can be adopted for metrics like \ac{WER} which are unbounded.
Alternatively, the cross-entropy training objective can be replaced by a Mean Squared Error (MSE) objective which does not require any bounded relevance assumption:
\begin{equation}
\mathcal{L}_{\mbox{\tiny{MSE}}}^{point}=\sum_{u=0}^{U-1}\sum_{i=0}^{M-1} {\| w_{u,i}-f(\bm{x}_{u,i})\|}^2.
\end{equation}

\subsubsection{Pair-wise training}
With point-wise training the model implicitly learns to rank the channels by learning to predict their absolute quality. However, it does not consider relative performance of the other channels. 
%One way to account for recordings from the same utterance is 
One way to account for the other microphones is to train the network in a siamese fashion, as it has been proposed in RankNet~\cite{Burges2005}. In this case, labels are not required to represent an absolute measure and thus even unbounded metrics can be used directly. 
For a given utterance $u$, let us consider feature vectors from two channels $x_{u,i}$ and $x_{u,j}$ with related relevance scores $w_{u,i}$ and $w_{u,j}$. We can define a binary pairwise label as:
\begin{equation}
y_{u,i,j}
\begin{cases}
1 & \text{if  }  w_{u,i} > w_{u,j},\\
0 & \text{otherwise}.
\end{cases}
\end{equation}
Note that $y_{u,i,j}$ is an hard label (i.e. either 1 or 0) whose value depends on which relevance $w_{u,i}, w_{u,j}$ is higher than the other, and thus on the relative ranking of the two channels.
For each training sample  $(\bm{x}_{u,i}, \bm{x}_{u,j},y_{u,i,j})$ we can then define a binary cross-entropy loss as:
\begin{equation}
\begin{split}
    \mathcal{L}_{u,i,j} & =y_{u,i,j}\log[P(w_{u,i}>w_{u,j})]\\
    & +(1-y_{u,i,j})\log[1-P(w_{u,i}>w_{u,j})],
\end{split}
\end{equation}
where $P(w_{u,i}>w_{u,j})$ is the probability estimated by the network $f(\cdot)$ that $\bm{x}_{u,i}$ is more relevant than $\bm{x}_{u,j}$, which can be computed as:
\begin{equation}
    P(w_{u,i}>w_{u,j}) = \sigma\left(f(\bm{x}_{u,i})-f(\bm{x}_{u,j})\right). 
\end{equation}
The overall training loss is obtained by summing over all unique microphone pairs and utterances:
\begin{equation}
\label{eq:ranknet}
    \mathcal{L}_{\mbox{\tiny{BCE}}}^{pair}=\sum_{u=0}^{U-1}\sum_{(i,j)\in \mathcal{I}_u} \mathcal{L}_{u,i,j}.
\end{equation}
where $\mathcal{I}_u=\left\{(i,j) : |w_{u,i}-w_{u,j}| > \delta, i \neq j \right\}$ is the set of microphone pairs whose relevance difference in utterance $u$ is larger than $\delta$ with $\delta \geq 0$.
% Note that in Eq.~\ref{eq:ranknet} we consider only pairs where one of the channels is more relevant than the other and discard 
%Note that all pairs where channels have the same or very similar relevance (in our case ~\ac{WA} or ~\ac{WER}) are discarded.
Thus the size of the training set is upper bounded to $(U-1)(M-1)(M-2)/2$.

\subsubsection{List-wise training}
In RankNet, the ranking network learns to order the items by comparing them only in a pairwise fashion. 
%
%However, in inference, this requires $\mathcal{O}(M^2)$ network forward passes as every possible pair should be compared in order to rank all channels. Thus, while effective, this approach can be computationally prohibitive for scenarios where a large number of microphones is employed. 
However, due to the use of hard labels, the learning process does not take into account the actual difference between two samples as it cares only for relative pair-wise ordering. 
Nonetheless, swapping the ranks of two samples with very similar relevance should be less critical than swapping two samples with a very different relevance.

%Although effective, this approach has one possible issues: the learning process is agnostic 
%of the actual difference between two samples and it may waste a lot of resources trying to properly order non relevant data. It is worth keeping in mind that, at the end, we are interested in the best channel. 
These problems can be addressed by  employing ListNet~\cite{Cao_listnet_2007}. Contrary to the pair-wise approach, for each utterance $u$ all available  microphones $M$ are used to compute a cross-entropy loss:
\begin{equation}
    \mathcal{L}_{\mbox{\tiny{XCE}}}^{list} = \sum_{u=0}^{U-1}\sum_{i=0}^M \mathcal{S}(w_{u,i})\log[\mathcal{S}(f(\bm{x}_{u,i}))].
\end{equation}
$\mathcal{S}(\cdot)$ is the softmax operator which ensures that both labels and network outputs can be treated as probability distributions. It also enforces that ranking, for each utterance, is determined only by relative performance of each microphone. %The total number of examples in ListNet is simply the number of utterances in the training set  $U-1$ as channels are considered all together in the loss.  

%\begin{figure*}[htbp]
%\centering
%\subfigure[]{
%\includegraphics[trim={1cm 5cm 1cm 5cm}, clip, width=.45\textwidth]{dev_wer}
%%}
%\subfigure[]{
%\includegraphics[trim={1cm 5cm 1cm 5cm}, clip,width=.45\textwidth]{eval_wer}
%}
%\caption{\ac{WER} on development and test sets using different model architectures and different training losses. Black lines refers to the RNN model while red lines to the TCN model.}
%\label{fig:wer_charts}
%\end{figure*}

\section{Experimental Analysis}
\label{sec:experiments}

\subsection{Datasets}
In order to evaluate our method we experimented with two datasets: a synthetic dataset generated on purpose and the data used in the CHi\-ME-6 challenge. We describe them thereafter. %For both datasets we employed the ASR back-end provided by the challenge organizers: the Kaldi recipes for CHiME-5~\cite{Barker2018} with the acoustic model presented in~\cite{Manohar_2019}. 

\subsubsection{Synthetic Dataset}
We generated a multi-channel synthetic dataset featuring an ad-hoc microphone network with 8 cardioid microphones. 
Clean speech utterances are uniformly sampled from LibriSpeech \cite{Panayotov-2015} using \texttt{train-clean-100} for training, \texttt{dev-clean} for validation and \texttt{test-clean} for test. 
We used a total of 20k utterances for train and 2k for validation and test splits. Point-source noise from the dataset in \cite{Furnon2020SE} is also employed to make the data more realistic. 
A different acoustic scenario is sampled for each utterance. Using gpuRIR \cite{diaz2021gpurir} we simulate a rectangular room whose size and  reverberation time (T60) are sampled uniformly between 10 and 60\,m2 and between 0.2 and 0.6\,s respectively. 
The positions and orientations of the speaker, noise and of the 8 microphones are chosen randomly inside the room but with the constraints that the speaker cannot be closer than 0.5\,m from any microphone or wall and each microphone should be at least 0.5\,m apart from any other. 
Relevance labels are obtained by training an ASR system on the training portion using a modified Kaldi \cite{povey2011kaldi} LibriSpeech recipe and computing the errors on such set. 

\subsubsection{CHi\-ME-6}
The CHi\-ME-6 Challenge~\cite{CHiME6} dataset features real dinner parties attended by 4 participants, recorded by 6 Kinect arrays, each with 4 microphones. Devices are distributed in space in order to cover the whole apartment, which may include multiple rooms. The dataset features also oracle speech segmentation and a manually selected reference device for each speech segment.
In our experiments we employed the ASR back-end provided by the challenge organizers using the official Kaldi recipe with the acoustic model and two-pass decoding strategy presented in~\cite{Manohar_2019}.

%MAYBE EXTEND? %In this work we used the official training, development and evaluation sets portions.

%We studied the proposed LTR framework with 2 different neural architectures: one composed of multiple stacked Recurrent Neural Network layers and a Temporal-Convolutional-Network (TCN). For each architecture we evaluated the 3 ranking losses described before in Section \ref{sec:learningtorank}. %We evaluated all combinations architecture-loss, resulting in 4 different systems.

\subsection{Neural Network Architecture}
We studied the proposed MicRank LTR framework with the Temporal-Convolutional-Network (TCN) used in \cite{cornell2020detecting} based on ConvTasNet separator \cite{luo2019conv}. We employ as input features 40 logmel filterbanks extracted from 25\,ms windows with 10\,ms stride. These are fed to a layer normalization and a $40 \times 64$ fully connected layer. This latter is followed by 3 blocks each comprised of 5 residual blocks with 1-D dilated convolutions. Each residual block has the same structure as described in \cite{cornell2020detecting}, with the dilation factor increasing for each successive residual block as $2^0, 2^1, \hdots, 2^{4}$. As in \cite{cornell2020detecting} we use 64 channels for bottleneck convolutions, 128 channels and a kernel size of 3 for depth-wise separable convolutional layers.

%The final layer is the same as described above for the recurrent model. We evaluate the performance of the model considering the 3 ranking losses described in Section \ref{sec:learningtorank}.
The network is fed a fixed-length input corresponding to 200 frames. Speech segments longer than 2\,s, are split in chunks which are processed individually. Zero-padding is used for segments shorter than 200 frames. During training, the same relevance is used for all chunks derived from the same speech segment. 
The network is applied to each channel independently and relevance is obtained via a final $40 \times 1$ fully connected layer followed by mean pooling over each 200 frames chunk. 
In inference, the output score is averaged over all chunks, which are extracted with an overlap factor of 4.
This architecture has a total of 266k parameters and is thus extremely light if compared with the AMs used in this work, making this approach significantly more computationally efficient than decoder and posterior-based techniques. 

Models are trained %on each dataset training set, 
using Stochastic Gradient Descent (SGD). To improve generalization, we employ SpecAugment \cite{park2019specaugment} based Mel-band masking. 
Learning rate, batch size, weight decay and SpecAugment parameters are tuned on each dataset %respective 
validation set. %We use \ac{WA} as the relevance score for each channel. \ac{WA} is computed by decoding and scoring each utterance and each channel of the training set using the ASR back-end. We experimented also with~\ac{WER} using normalization strategies but we did not observe noticeable differences. 
As relevance score we use \ac{WA}, which is computed by %decoding and 
scoring each utterance and each channel of the training set using the ASR back-end. We experimented also with~\ac{WER} using normalization strategies but we did not observe noticeable differences.
\subsection{Oracle and Baseline Methods}
\label{sec:comp-methods}
%We compare our proposed approach against the CHiME6 baseline, an oracle upper bound and the channel selection based on envelope variance presented in~\cite{WOLF_EV_2014}. For all methods we use the same ASR backend. Note that, as per the related Kaldi recipe, the insertion penalty and language model weight are optimized for each method on the development set. 

We evaluate our proposed method against a set of baselines and oracle approaches. We consider as the upper-bound for this task the oracle channel selection obtained by taking, for each utterance, the channel with lowest~\ac{WER} among all the available ones. Moreover, we consider a set of selection strategies that relies on the distance between the speaker and microphones and on oracle signal-based quality metrics. Regarding the latter, we consider Short-Time Objective Intelligibility STOI \cite{jensen2016algorithm}, 
Signal-to-Distortion Ratio (SDR) \cite{vincent2006performance} and Perceptual Evaluation of Speech Quality (PESQ) \cite{rix2001perceptual}. These are computed with respect to oracle non-reverberated clean speech for the synthetic dataset and with respect to close-talk per-speaker microphones for CHi\-ME-6.

As oracle distance from the speaker is not available in CHi\-ME-6, we instead consider the baseline system provided by the challenge organizers which employs ~\ac{WPE}~\cite{WPE} followed by BeamformIt~\cite{Beamformit} applied on a ``pseudo-oracle" manually selected array for each utterance. The manual selection is based on the positions and orientations of the speakers obtained via video recordings and is provided by challenge organizers. The alternative, more performing, baseline system based on Guided Source Separation \cite{boeddeker2018front} is not considered here as it also exploit oracle diarization. 

In addition, we evaluate MicRank against three aforementioned state-of-the-art channel selection methods.
In detail, we consider the posterior-based approach proposed in \cite{Xiong2018} (AM-Entropy in the following) for the synthetic dataset only and two signal-based approaches for both datasets EV~\cite{WOLF_EV_2014} and CD~\cite{Guerrero2018}. For the former we used the pre-trained LibriSpeech AM available in Kaldi. 
Regarding \cite{Guerrero2018} we evaluate both the blind version (CD-blind) as well as the oracle version (CD-informed) computed in the same way as the aforementioned signal-based oracle measures. 
Sub-band weights in EV are tuned on each dataset training set using SGD and a cross-entropy objective for selecting the best channel. 

\vspace{-0.1cm}
\section{Results}\label{sec:results}

\begin{figure}
\centering
\includegraphics[width=8.5cm]{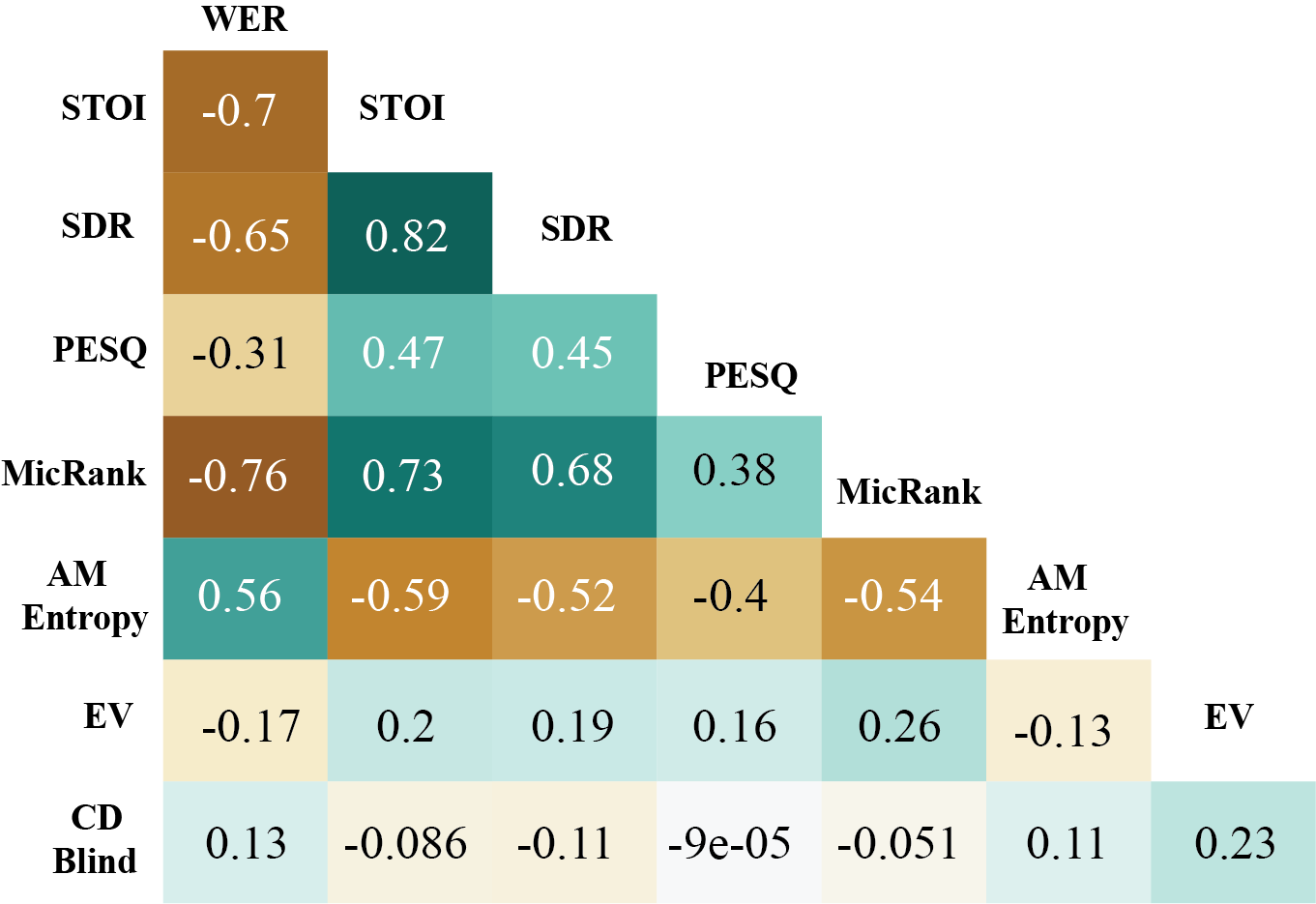}
\caption{Pearson correlation plot for different channel selection techniques on synthetic data. MicRank refers to the ListNet-based method here.}
\label{fig:corrplot}
\vspace{-0.5cm}
\end{figure}

Results on the synthetic dataset are reported in Table~\ref{tab:result_synth}. The upper part of the Table reports results obtained by randomly selecting one of the microphones as well as using oracle measures. Note that, as expected, picking the closest microphone leads to better~\ac{WER} with respect to a random choice. Nevertheless, this is not the best strategy as signal-based oracles further improves the performance with STOI providing the best results. %Nevertheless, the~\ac{WER} achievable by an oracle channel selection based on the~\ac{WER} itself is still out of reach.
Among blind channel selection techniques, EV and AM-Entropy considerably improve over random selection and perform slightly worse than the oracles.
All MicRank-based techniques are able to bring substantial gains over such previous blind selection methods. In particular, we can observe that, as expected, pair-wise and list-wise methods outperform point-wise ones which cannot account for relative performance. Notably, the best WER for RankNet and ListNet is lower than the Top-3 averaged WER of oracle WER selection, indicating that these methods are able to pick up always the best or second-best channel among the top 3. 
Amidst previously proposed selection methods, EV and AM-Entropy have comparable performance despite the former is remarkably less computational expensive.

\begin{table}[!htbp]
\vspace{-0.2cm}
    \centering
    \setlength{\tabcolsep}{7pt} % Default value: 6pt
    \renewcommand{\arraystretch}{1} % Default value: 1
    \footnotesize
    \begin{tabular}{|cl|cc|cc|}

    \multicolumn{6}{c}{\normalsize Synthetic Dataset}\\
    \hline
         \multicolumn{2}{|c|}{{ \bf Ranking Method}} & \multicolumn{2}{c}{{ \bf Dev}} &  \multicolumn{2}{c|}{{ \bf Test}}\\
        
         & & Best & Top-3 & Best & Top-3 \\
         \hline
        \multicolumn{2}{|r|}{Random Selection} & 51.7 & 51.5 & 40.9  & 41.1 \\
        \hline
                \multirow{6}{*}{oracle} & CD-Informed~\cite{Guerrero2018} & 45.1  & 47.7 & 36.9  & 38.3 \\
                                &  PESQ & 41.9  & 45.8 & 33.1  & 36.4 \\
                                & closest & 37.0  & 45.1 & 29.9  & 36.1\\
        %reverberant speaker utterance energy &0.579&	0.466\\	 
                                &  SDR  & 37.4  & 43.8 & 29.6 & 34.9 \\
                                &  STOI & 36.3  & 44.2  & 29.2 & 35.2 \\
                                &  WER  & 32.0 & 39.6 &  24.8 &  30.6 \\
        \hline
        \hline
         \multirow{3}{*}{baseline} & CD-blind~\cite{Guerrero2018} & 46.1 & 48.1 & 36.2  & 39.4 \\
         & EV~\cite{WOLF_EV_2014} & 39.0  & 44.9 & 31.8 & 35.8 \\
         & AM-Entropy~\cite{Xiong2018}  & 41.2  &  45.8 &  31.1  & 35.5 \\

        \hline
        \hline
        \multirow{4}{*}{MicRank}& Point-wise XCE & 37.3 & 44.1   & 30.4 & 34.6 \\
                            & Point-wise MSE & 36.9 & 43.7   & 30.0 & 34.3 \\
                            & RankNet   & 36.5 & 43.4 & 28.8 & 34.1\\
                            & ListNet   & \bf{36.0} & \bf{43.2}  & \bf{28.5} & \bf{33.9} \\
        \hline
    \end{tabular}
    %\caption{\ac{WER} on the development and evaluation sets of the synthetic dataset. We report both the best WER as well as the average WER obtained by Top-3 selected microphones.}
    \caption{\ac{WER} on the synthetic dataset. We report both the best WER as well as the average WER on the Top-3 selected microphones.}
    \label{tab:result_synth}
    \vspace{-0.2cm}
\end{table}

In Figure \ref{fig:corrplot} we report a Pearson correlation plot for a subset of selection metrics obtained on synthetic dataset test set. 
Interestingly EV obtains rather low correlation with WER despite its efficacy in selecting favorable channels as shown in Table \ref{tab:result_synth}. This is because we found that EV is able to pick favourable channels but fails to correlate with WER for the unfavourable ones. 
CD-Blind has the same behaviour while AM-Entropy, which is posterior based, shows much better correlation even for unfavourable channels.
Again, we can notice that the proposed method is the one with highest absolute correlation value and surpasses even some oracle measures. 

%The lower part of the table analizes the results provided by different flavours of our proposed MicRank framework. First of all, note that all strategies outperform the state-of-the-art performance of EV. In addition, the best of the proposed approaches perform slightly better that the signal-based oracles, meaning that MicRank is capable of identifying what features the ASR acoustic model likes the most. Between the various MicRank approaches, xxx is the best performing on this dataset.
%First of all we analyze the impact of using RankNet or ListNet with respect to a point-wise loss. Figure~\ref{fig:wer_charts} reports the \ac{WER} on both development and test sets using two neural architectures with and without ranking losses. The~\ac{WER} obtained by randomly selecting the channels is approximately 72.5\% on the development set and 67.0\% on the test set. The figure shows that networks trained with the point-wise loss are unable to learn how to properly score the channels. Among the two ranking-based losses, RankNet seems slightly better than ListNet, although the performance is very similar. When trained with RankNet, both architectures deliver very similar performance.

\begin{table}[!htbp]
    \vspace{-0.5cm}
    \centering
    \setlength{\tabcolsep}{7pt} % Default value: 6pt
    \renewcommand{\arraystretch}{1} % Default value: 1
    \footnotesize
    \begin{tabular}{|cl|c|c|}

    \multicolumn{4}{c}{\normalsize CHi\-ME-6 Dataset}\\
    \hline
         \multicolumn{2}{|c|}{{ \bf Ranking Method}} & { \bf Dev} &  { \bf Eval}\\
        
         & & Best &  Best \\
         \hline
        & Random Selection & 73.1  &  68.0 \\
        \hline
                \multirow{6}{*}{oracle} & CD-Informed~\cite{Guerrero2018} & 70.8 &  68.7    \\
                                &  PESQ & 66.0  & 60.1   \\
        %reverberant speaker utterance energy &0.579&	0.466\\	 
                                &  SDR  & 65.2 & 58.9   \\
                                &  STOI & 64.8  & 58.5  \\
                                &  WER  & 56.7 & 51.3    \\
                                &  CHi\-ME-6 Baseline & 69.2  &  60.5  \\
        \hline
        \hline
         \multirow{2}{*}{baseline} & CD-blind~\cite{Guerrero2018} & 72.5 &  67.0   \\
         & EV~\cite{WOLF_EV_2014} & 68.6  & 59.9   \\

        \hline

        \hline
        \hline
        \multirow{2}{*}{MicRank}
                            & RankNet   & 67.4  & \bf{59.0}  \\
                            & ListNet   & \bf{67.2} & 59.5 \\
                            %\hline
                            %& ListNet-Synth   & 68.5 & 61.5 \\
        \hline
    \end{tabular}
    \caption{\ac{WER} on CHi\-ME-6 development and evaluation sets.}
    \label{tab:chime6_results}
    \vspace{-0.4cm}
\end{table}

%\begin{table}[!htbp]
%    \centering
%    \begin{tabular}{|c|c|c|c|}
%%   \hline
%         Model & Loss & Dev & Eval \\
%         \hline
%        \multirow{ 2}{*}{TCN} & ListNet& 67.17 & 59.56\\
%        & RankNet& 67.41 & 59.04\\
%        \hline
%        \multirow{ 2}{*}{RNN} & ListNet & 67.81 & 60.73\\
%        &RankNet & 67.11 & 59.31\\
%        \hline
%    \end{tabular}
%    \caption{\ac{WER} obtained with the proposed neural channel selection approach with %different models and training strategies. The best model is selected on the development set.}
%%    \label{tab:Wer_NN}
%\end{table}

Finally, in Table~\ref{tab:chime6_results} we report the performance achieved on CHi\-ME-6 data for the most promising approaches as found on the synthetic set. Note that both EV and MicRank methods considerably improve with respect to the CH\-iME-6 Baseline, which benefits from ``pseudo-oracle" knowledge of the speaker position and features dereverberation plus beamforming. Both RankNet and ListNet based systems improve over EV but, contrary to the synthetic dataset, are unable to outperform signal-based oracle-level performance especially on the development set. This is mainly due to the fact that CHi\-ME-6 features a substantial amount of overlapped speech \cite{cornell2020detecting}, while in the synthetic data only one speaker is present. This occurs in particular in the development set, which is where we observe the largest difference between signal-based oracles and the proposed method. Current selection methods, including MicRank, are unable to account for speaker identity when ranking the channels for a given utterance. This can lead to mistakenly rank the channels with respect to the interfering speaker. On the other hand, signal-based oracle measures are able to implicitly account for this because they are computed with respect to the correct speaker close-talk microphone. RankNet seems to generalize better than ListNet on CHiME-6 due to the fact that on CHiME-6 relevances are very close to each other in the training set but not in the dev and eval sets, and thus using an hard label, as in RankNet, can help boosting discriminability and generalization. 

%Nevertheless also here we can observe that the pair-wise and point-wise approaches outperform the point-wise ones. RankNet seems to generalize better than ListNet on CHiME-6 due to the fact that on CHiME-6 training relevances are very close to each other and thus using an hard label, as in RankNet, can help boosting discriminability. 
%In general, we can see that on both datasets pair-wise and list-wise methods outperform point-wise approaches but 
% 
% While on synthetic dataset ListNet was 

%In general the two channel selection approaches (\cite{WOLF_EV_2014} and ours) are better than the baseline. Finally, the best configuration for our proposed method (TCN architecture trained with RankNet) slightly improves the performance of the selection based on~\ac{EV}.

\section{Conclusions}
\label{sec:conclusions}
In this paper we proposed MicRank, a fully neural channel selection framework for ad-hoc microphone arrays. In this framework the channel selection problem is formulated as a learning to rank (LTR) problem and a DNN is trained to rank the microphones using directly ASR errors on a training set. 
We explored three different LTR training strategies and validated our method on a synthetic dataset and CHi\-ME-6. 
We showed that the proposed method is able to outperform 
previous state-of-the-art channel selection approaches which rely on signal-based or posterior-based features and is even able to surpass oracle signal-based selection on single-speaker synthetic data. 
Besides investigating other LTR training strategies in further work we could explore how to condition the channel selection on speaker identity in order to improve the performance on multi-party scenarios such as CHi\-ME-6. Moreover, it would be interesting to study how much performance changes if a different back-end ASR from the one used in testing is used to generate the relevance labels.
%Another issue to study is the impact of the specific back-end used to generate the relevance labels and the relationship with the decoder at hand.

%We evaluated two neural architectures and three different ranking training strategies on the data used in the CHi\-ME-5 and CHi\-ME-6 challenges. We obtained promising results which show that the networks learn, to a certain extent, how to select good channels. This allows us to reduce the~\ac{WER} with respect to the baseline without any speech enhancement technique and in a completely agnostic way with respect to the recording set up.

%Besides investigating other architectures and hyper-parameters, one open issue in implementing learning-to-rank frameworks in ASR contexts is how to incorporate more advanced ranking metrics in the training loss, such as LambdaRank~\cite{Yue2007OnUS}.

\section{Acknowledgements}
%The research reported here started at the 2019 Frederick Jelinek Memorial Summer Workshop on Speech and Language Technologies, hosted at L'\'Ecole de Technologie Sup\'erieure (Montreal, Canada) and sponsored by Johns Hopkins University with unrestricted gifts from Amazon, Facebook, Google, and Microsoft. The authors would like to thank Maurizio Omologo for his valuable contribution to this research.
The research reported here started at JSALT 2019, hosted at ETS (Montreal, Canada) and sponsored by JHU with gifts from Amazon, Facebook, Google, and Microsoft. The authors would like to thank Maurizio Omologo for his valuable contribution to this research.
\newpage
\bibliographystyle{IEEEbib}
\bibliography{mybib,conferences,IEEEfull}

\end{document}